\documentclass[letterpaper, 12pt]{article}

\usepackage{amsmath, bbm}
\usepackage{amssymb}
\usepackage{mathptmx}
\usepackage{graphics}
\usepackage{graphicx}
\usepackage[top=1.25in, bottom=1.25in, left=1in, right=1in]{geometry}

\newcommand {\beq}{\begin{equation}}
\newcommand {\eeq}{\end{equation}}
\newcommand {\bea}{\begin{eqnarray}}
\newcommand {\eea}{\end{eqnarray}}
\newcommand{\bo}{\raise-.5mm\hbox{\large$\Box$}} 

\begin{document}
\begin{titlepage}
\begin{flushright}
YITP-SB-06-01 \\ ITFA-2006-09 \\
\end{flushright}
\vskip 22mm
\begin{center}
{\LARGE {\bf Does Smoothing Matter?}}
\vskip .25in
{\bf Martin Ro\v{c}ek}\\
{\em C.N. Yang Institute for Theoretical Physics \\
SUNY, Stony Brook, NY 11794-3840, USA}\\ 
{\tt Martin.Rocek@stonybrook.edu}
\vskip .25in
{\bf Patrick van Nieuwenhuizen}\\
{\tt patvann@gmail.com}
\vskip .25in
{\bf Abstract} \end{center}
\begin{quote}
We study how inhomogeneities modify 
large scale parameters in General Relativity. 
For a particular model, we obtain exact results: 
we compare an infinite string of extremal black holes 
to a corresponding smooth line with the same mass and 
charge in five dimensions. We find that the effective energy 
density does not differ significantly.
\end{quote}
\vfill
\end{titlepage}

\section{Introduction}
Less than $5\%$ of all matter and energy in the universe 
is visible--the origin of the rest is a mystery.  
Dark Energy (DE) constitutes perhaps $70\%$ of this missing mass 
(see, {\it e.g.}, \cite{DE}); this is the energy needed to accelerate 
the whole universe outwards at its measured rate.  
The remaining $25\sim30\%$ is Dark Matter (DM) 
(see, {\it e.g.}, \cite{BHS}), and is needed
to account for the proper motions of celestial bodies.

This paper investigates how
averaging out inhomogeneities in matter affects the 
measurement of mass and energy parameters.
When we calculate interactions between celestial bodies, 
we generally use a smooth mass density profile in galaxies or clusters;
that is, we average mass distributions to simplify our calculations.  
However, Einstein's theory is nonlinear, and this averaging 
could potentially alter what energies we expect to see.  
The averaged stress tensor of the real metric need 
not be the same as the real stress tensor of the averaged metric;
this discrepancy could be pertinent to the problem of DM.

Effects of inhomogeneities and DM seem to have some 
characteristics in common.  Smoothing of mass fluctuations 
should become less important at greater 
distances from the fluctuations; thus the effects on 
energy density would be greater closer to the masses, which 
is where halos are identified. It also makes sense 
that the existence of halos is contingent upon some 
intrinsic properties of mass distributions in general, because 
the halos are found on such different scales 
(galaxies, galaxy clusters, even superclusters).  
This marks a difference from DE, which is seen as a constant 
energy density throughout the whole universe, a 
``cosmological constant'' as Einstein originally called it.

This work ties into the ``fitting problem" first described by George Ellis in 1983 \cite{ellis}. Ellis discussed the problem of how to average out inhomogeneities with respect to cosmological models of the universe, and thus, from a modern perspective, his work has more relevance to DE than DM. The effect of inhomogeneities for cosmology has also been called ``back-reaction." There has been significant discussion of back-reaction and DE: see, {\it e.g.}, \cite{other}. Some of these works claim that back-reaction could account for a cosmological constant and obviate the need for DE; others refute this claim.
  
In the context of DM, F.~Cooperstock and S.~Tieu claimed to construct
a model in which a galaxy's need for a halo was canceled by 
including some of its possible rotational energy \cite{cern}; 
M.~Korzynski and others have argued convincingly that this claim is erroneous \cite{korz}.
  
All these calculations, however, are perturbative--based upon approximations; 
until now, there has not been an {\em exact} calculation 
of the impact of smoothing on energy.\footnote{We thank 
S. Rasanen for making us aware of exact cosmological solutions in 
a specific model with inhomogeneous dust distributions considered by Kozaki and Nakao
in \cite{other}.} When perturbative calculations do not lead to large effects, one can
take them seriously; however, computations that claim large effects in perturbation
theory are by their nature uncontrolled approximations and cannot be trusted without an
exact treatment. 
This paper determines the exact energy for two scenarios, 
one of which is a `smoothed' version of the other. Both
scenarios use BPS black hole solutions because
it is easy to find a solution for any sum of these special 
black holes, as the electromagnetic repulsions and 
gravitational attractions between them cancel \cite{maj}.  
Specifically, we compare an infinite string of discrete 
charged black holes to a smooth
line of charge, with both distributions normalized to
have the same average charge per unit length.
We work in five dimensions because
such distributions are divergent in four dimensions. We find that in our specific
calculation, inhomogeneities cannot give rise to apparent DM.

\section{The Metrics}
\setcounter{equation}{0}
The solution of the coupled Einstein-Maxwell field equations 
which describes a set of $n=1\dots N$ extremal black holes with 
charges $q_n$ in five dimensions\footnote{In four dimensions,
the solution \cite{maj,gkltt} has the form 
$ds^2=-{\Omega}^{-2}dt^2+{\Omega}^2d{\vec x}^2,~
A_{0}={\Omega}^{-1}$.} is \cite{gkltt}:
\beq \label{(4.1)}
ds^2=-{\Omega}^{-2}dt^2+{\Omega}{\ }d{\vec x}^2~,
\qquad A_{0}=\frac{\sqrt 3}2{\Omega}^{-1}~,\qquad
{\Omega}=1+\sum_{n=1}^N \frac{q_n}{{|{\vec x}-{\vec x}_n|^2}}~~; 
\eeq
as we are in five dimensions, $d{\vec x}^2=dx^2+dy^2+dz^2+dv^2$.  
We note that $\Omega$ is time-independent and harmonic 
(it satisfies ${\nabla}^2{\Omega}=0$ outside the source), and that, in these coordinates,
the horizon of each black hole is shrunk down to a point.

These are BPS solutions, with the property that gravitational 
attraction and electromagnetic repulsion between any set of them cancel \cite{maj}. 
This makes them time-independent, and allows for the simple solution in (\ref{(4.1)}).  

We consider two configurations: a discrete and a smooth 
solution. The discrete solution is for a set of black 
holes positioned along a line at equal distances $L$ from one 
another. The smooth solution is for a smooth charged string along an axis. 
The continuous set can be viewed as a 
limiting case of the discrete set, namely the limit as the number of 
black holes per unit length tends to infinity and as the charge of each 
black hole tends towards zero, in such a way that the total charge
per unit length remains the same.  

\subsection{The Discrete Solution}

For the discrete case, we choose black holes with the same charge $q_D$;
thus $\Omega_D$ is given by:
\beq \label{(4.2)} 
\Omega_D=1+\sum_{n=-\infty}^{\infty} \frac{q_D}{r^2+(v-n)^2}~,
\eeq
where black holes are one unit length apart on the 
$v$-axis and $r$ is the distance to the $v$-axis. We can restore the
separation $L$ by dimensional analysis 
(keeping the total charge per unit length constant):
\beq \label{(4.3)} 
\Omega_D^L=1+\sum_{n=-\infty}^{\infty} \frac{Lq_D}{r^2+(v-nL)^2}
=1+\sum_{n=-\infty}^{\infty} \frac{q_D/L}{(r/L)^2+(n-[v/L])^2}~.
\eeq
Thus we can start with $L=1$ and find the dependence on $L$ simply by
making the substitution $(r,v,q_D)\to (r/L,v/L,q_D/L)$. 

To evaluate the sum in (\ref{(4.2)}), we use contour integration. We find
\beq 
\sum_{n=-\infty}^{\infty}\,{\frac{1}{r^2+(v-n)^2}}
=\frac{i{\pi}{\cot}(\pi(v+ir))}{2r}-\frac{i{\pi}{\cot}(\pi(v-ir))}{2r}~~,
\eeq
which simplifies to
\beq\label{(4.9)}
\sum_{n=-\infty}^{\infty}\, \frac{1}{r^2+(v-n)^2}
={\frac{\pi}{r}}{\frac{{\sinh}(2{\pi}r)}{{\cosh}(2{\pi}r)-{\cos}(2{\pi}v)}}~~.
\eeq
Thus, using (\ref{(4.2)}) and (\ref{(4.9)}), the metric for 
the discrete solution is given by:
\beq \label{omd}
\Omega_{D}=1+{\frac{{\pi}q_D}{r}}
{\frac{{\sinh}(2{\pi}r)}{{\cosh}(2{\pi}r)-{\cos}(2{\pi}v)}}~~.
\eeq
\subsection{The Smooth Solution}
The metric for the smooth configuration
is the same as the limit of the metric of the discrete configuration
as the masses and the distances between them approach zero.  
From (\ref{(4.3)}), we find:
\beq \label{oms}
\Omega_{S}=1+\lim_{L\to0}~{\sum_{n
=-\infty}^{\infty} \,\frac{L\,q_D}{r^2+(v-nL)^2}}
=1+\int_{-{\infty}}^{\infty}dx~{\frac{q_D}{r^2+(v-x)^2}}
=1+\frac{{\pi}q_D}{r} ~.
\eeq
As expected, the discrete harmonic function $\Omega_D$ (\ref{omd}) approaches the smooth 
function $\Omega_S$ (\ref{oms}) when the distance $r$ to the $v$-axis is large.
\section{Energy}
\setcounter{equation}{0}
Since we have normalized the two solutions 
so that they have the same average charge density along the $v$ axis,
and since the solutions are BPS, they must have the same total mass density
as well; this does not immediately rule out a possible effect, as we discuss below.

We check this using Weinberg's definition of total energy \cite{weinberg}:  
we rewrite the metric as a flat metric plus a correction term $h$:
$g_{\mu\nu}=\eta_{\mu\nu}+h_{\mu\nu}$.  
We split the Einstein tensor into terms linear and non-linear in $h$: 
$G_{\mu\nu}^{(L)}(h)+G_{\mu\nu}^{(NL)}(h)=T_{\mu\nu}$.  
Then the definition of the total (matter plus gravitational) energy density of a 
system is given by \cite{weinberg}:
\beq \label{(2.4)} 
\mbox{Total Energy Density}=G_{00}^{(L)}(h)=T_{00}-G_{00}^{(NL)}(h) ~.
\eeq
We integrate $G_{00}^{(L)}(h)$ 
over all three directions ($x,y,z$) orthogonal to the line of black holes. 
Integrating along the entire $v$-axis would make the 
integral divergent--but this is not a problem: since the smooth
configuration is independent 
of $v$, and the discrete configuration
is periodically dependent upon it, we 
need only integrate over one period of the discrete configuration
({\it i.e.}, from 0 to 1 for $L=1$).
To perform the integral, we transform to 4-d cylindrical coordinates ($r$, $\theta$, $\varphi$, 
and $v$), and integrate over $v,\theta,\varphi$, factoring in the Jacobian $r^2\sin (\varphi)$
for the coordinate change. Then the total energy per unit length along the $v$-axis (describing an
infinite slice of space orthogonal to the $v$-axis, with a width of one unit length) can be written as
the following surface integral:
\beq \label{totE}
E=\int_0^1 dv \int d\theta d\varphi \, r^2\sin (\varphi)\, 
n^i (\partial_i h_{jj} - \partial_j h_{ij})   ~,
\eeq
where $\vec n$ is a unit vector normal to the $v$-axis. Since at large
values of $r$ the integrands are the same, 
this integral yields the same result for our two cases--in 
agreement with the requirements of the BPS condition 
(which ensures that interaction energies cancel) and the 
equality of charge per unit length (which ensures that 
the energies which are not due to interactions are also equal).

The equality of the total masses does not in itself 
preclude the possibility of an effect on DM: the energy which enters Einstein's 
equations, and so is relevant to DM, is the covariant 
matter energy $T_{00}$. This quantity is not necessarily 
the same for our two different distributions of black holes; comparing them
shows us how matter energy is affected by smoothing.

\subsection{Stress Tensor and Integration}

The matter energy of a system is given by the integral of $T_{00}$ 
over all space.  In a general spacetime with a time-like Killing
vector $\xi^\nu$ the following total derivative vanishes:
\beq
0=\int d^5x~\sqrt{-g}\,D_\rho \!\left(g^{\rho\mu}T_{\mu\nu}\xi^\nu\right)
=\int d^5x~\partial_\rho\!\left(\sqrt{-g}g^{\rho\mu}T_{\mu\nu}\xi^\nu\right)~;
\eeq
consequently, the contribution on a spacelike slice defines a conserved
energy. In our case, since both metrics are static and diagonal, 
the total matter energy reduces to:
\beq 
E=-\int{d^4x~\sqrt{-g}}\,g^{00}\,T_{00}~~.
\eeq
The stress-tensor is usually defined with a factor $\frac1{4\pi}$, but as
we only study ratios of energies, we are not interested in the overall normalization.

We calculate $T_{00}$ using
\beq 
T_{\mu\nu}=F_{\mu\alpha}F_{\nu\beta}g^{\alpha\beta}-{\frac14}
g_{\mu\nu}(F_{\alpha\beta}F_{\gamma\delta}g^{\alpha\gamma}
g^{\beta\delta})~,\qquad F_{\mu\nu}
={\partial}_{\mu}A_{\nu}-{\partial}_{\nu}A_{\mu}~, 
\eeq
where $A$ is the gauge potential (\ref{(4.1)}). Once again, because the
metrics are static and diagonal, and furthermore, because the gauge potential $A$ is purely
timelike and static, this simplifies, and we find
\beq 
T_{00}=F_{0j}F_{0j}g^{jj}-\frac{1}{4}g_{00}(2F_{0j}F_{0j}g^{00}g^{jj})~. 
\eeq
Recalling $A_{0}=\frac{\sqrt 3}2{\Omega}^{-1}$, $g_{00}=-{\Omega}^{-2}$, 
$g_{jj}={\Omega}$, and the determinant 
$g=- \Omega^2$ from equation (\ref{(4.1)}), we find:
\beq \label{(5.10)}
E=-\int d^4x~\sqrt{-g}\,g^{00}\,T_{00}
=6\int d^4x\left(\frac{\partial_{j}\Omega}{\Omega}\right)^2~.
\eeq
Here $(\partial_{j}\Omega)^2
=(\partial_{x}\Omega)^2+({\partial}_{y}\Omega)^2
+({\partial}_{z}\Omega)^2+({\partial}_{v}\Omega)^2$.

\subsection{Exact Energies}
As discussed above  equation (\ref{totE}), we integrate stress-energy 
over all three directions $(x,y,z)$ orthogonal to 
the line of black holes and integrate over one 
period of the discrete configuration over the $v$-axis. 
The smooth configuration
is the simpler of the two; we calculate it first. As discussed in section 2.1, we use $L=1$, and 
reintroduce $L$ later by rescaling.  We ignore integration over 
$\theta$ and $\varphi$ in both configurations because it only produces 
equal overall constants; we also drop the $\sin (\varphi)$ factor from the
Jacobian. Defining $k=\pi q$ and
$\Omega_S=1+\frac{k}{r}$, we find $(\partial_{j}\Omega_S)^2=k^2/r^4$, and then
(\ref{(5.10)}) gives
\beq \label{3.6} 
E_S=\int_0^\infty dr~{\frac{k^2}{(r+k)^2}}=k~.
\eeq
For the discrete configuration, the calculation is more complicated. 
We define $s=\sinh(2\pi r)$, $c=\cosh(2\pi r)$, $t=\sin(2\pi v)$, 
and $d=\cos(2\pi v)$ and compute:
\beq 
(\partial_{j}{\Omega}_D)^2={\frac{k^2}{r^4}}
{\frac{[s(d-c)+2\pi r(1-cd)]^2+(2\pi rst)^2}{(c-d)^4}} ~.
\eeq
Then, with $\Omega_D
=1+{\frac{k}{r}}{\frac{s}{c-d}}$, and the measure factor $r^2$, the 
energy for the discrete configuration is given by:
\beq \label{3.8} 
E_D=\int_0^1 dv\int_0^\infty \!dr~
k^2\frac{[s(d-c)+2\pi r(1-cd)]^2+(2\pi rst)^2}
{(c-d)^2(r(c-d)+ks)^2} ~.
\eeq
This is not easily evaluated as an indefinite integral, but we can 
compare $E_S$ and $E_D$ numerically. To understand the results, 
let us clarify what we should test. For $L=1$ and various 
$k$ (charge), we want to find the differences in energy between the two configurations. However, a smaller $k$ will certainly yield a 
smaller energy and smaller energy discrepancy, so to compare 
various $k$'s we divide by $E_S$ because $k \propto E_S$.

Thus we study the ratio
$\left(\int_0^\infty\int_0^1(E_D-E_S)\,dv\,dr\right)/\left(
\int_0^\infty\int_0^1E_S\,dv\,dr\right)$ as a 
function of $k$. Different $k$ represent different densities, as we 
have scaled the separation $L$ to $L=1$.  The results, computed 
with Maple, are consistent with zero; the actual numerical integrations
are somewhat delicate because of divergences in the integrand ($E_D$ diverges
when $r=0$ and $v=0,1$).

\section{Physical Application}

Though we did not find an effect, we discuss how any effect that we might
have found could have been interpreted physically.
The first step would have been to choose a scale, because DM is 
found on several different levels: galaxies, galaxy clusters, superclusters.  
Then we would determine which objects are smoothed for the situation chosen.
For example, to see how smoothing might create a halo around a galaxy, 
we would smooth the stars inside the galaxies, as is the common practice when calculating intergalactic motions.  

Our numerical results were calculated for $L=1$.  We can use these 
results, however, if we rescale the situation by changing the mass so 
that the density of the galaxy is correct. The 
value of $k$ in equations (\ref{3.6}) to (\ref{3.8}) would then be 
modified to provide the correct mass matching up 
with the real-life situation.  The $k$ we used was $\pi q$, and 
$q=Gm$ for extremal black holes.  The ratio to demand for 
similarity is $k/1=\pi Gm/L$.

So, we need the mass ($m$) and average distance between masses ($L$) as input, 
and then we can see what the energy difference is for $k=\pi Gm/L$.  
Let us apply these ideas to the situation of galaxy halos just discussed, testing 
how the smoothing of stars within the galaxies could affect the energy.  
We use planck-units, so that $G=1$.  The average distance between 
adjacent stars is about 10 light-years, $10^{52}$ Planck lengths.  
The average mass of a star is a bit less than $10^{39}$ Planck masses.  
Our $k$ is then around $10^{-13}$.  

Next let us apply the smoothing to galaxies within galaxy clusters.  
The average galaxy mass is $10^{45}$ Planck-masses.  
The average separation between galaxies is $10^7$ light-years, 
$10^{58}$ Planck-lengths.  So $k$ for this situation is 
again $10^{-13}$.

A third scale at which to look is that of superclusters.  
Superclusters consist of about a dozen clusters, each 
with a mass of about of $10^{55}$ Planck-masses.
Usually they are arranged in strings with lengths of 
10-100 Mpc, $10^{58}$ or $10^{59}$ Planck-lengths, 
so the distance between each is about $10^{57}$ or $10^{58}$.  
Our $k$ is then $10^{-2}$ or $10^{-3}$.  

\section{Conclusion}

The calculation performed in this paper is by no means flawless.  
Five-dimensional effects are expected to show similar trends 
as their four-dimensional counterparts, but when one is discussing 
the extent of an effect, 5-d is less applicable.  
Also, these calculations were done with BPS solutions.  
They are easy to work with but certainly do not represent 
standard matter.  They are maximally charged and describe point-like masses,
which is of course physically unrealistic.

These are the first exact calculations of the effects of 
inhomogeneities on the apparent mass of astrophysical objects; as such, they complement and confirm 
perturbative arguments.  There is room for development--our 
calculation was based on a particular model, and calculations in other models could help 
elucidate the exact effects of smoothing.

\vskip .5cm
\noindent{\bf Acknowledgements}

\noindent 
We are happy to thank Peter van Nieuwenhuizen for extensive discussions, encouragement,
and comments. We also thank Nick Jones for comments, and numerous authors for making
us aware of the literature after our paper was posted on the arXiv.
MR thanks the Institute for Theoretical Physics  
at the University of Amsterdam for hospitality. MR was supported in part by
NSF grant no.~PHY-0354776, by the University of Amsterdam, and by Stichting FOM.

\end{document}